\documentclass[epj]{svjour}
\usepackage{graphics}
\newcommand{\bq}{\mbox{\boldmath $q$}}
\newcommand{\bp}{\mbox{\boldmath $p$}}
\newcommand{\bk}{\mbox{\boldmath $k$}}
\newcommand{\br}{\mbox{\boldmath $r$}}

\newcommand{\bx}{\mbox{\boldmath $x$}}
\newcommand{\by}{\mbox{\boldmath $y$}}

\newcommand{\bL}{\mbox{\boldmath $L$}}

\newcommand{\bS}{\mbox{\boldmath $S$}}
\newcommand{\bR}{\mbox{\boldmath $R$}}
\newcommand{\bZ}{\mbox{\boldmath $Z$}}
\newcommand{\bsigma}{\mbox{\boldmath $\sigma$}}
\newcommand{\btau}{\mbox{\boldmath $\tau$}}

\newcommand{\bxs}{\mbox{\scriptsize \boldmath $x$}}

\newcommand{\bqs}{\mbox{\scriptsize \boldmath $q$}}
\newcommand{\bks}{\mbox{\scriptsize \boldmath $k$}}

\newcommand{\bZs}{\mbox{\scriptsize\boldmath $Z$}}

\newcommand{\beqa}{\begin{eqnarray}}
\newcommand{\eeqa}{\end{eqnarray}}
\newcommand{\nn}{\nonumber}

\begin{document}

\title{
Nucleon-nucleon interactions via Lattice QCD: Methodology}
\subtitle{HAL QCD approach to extract hadronic interactions in lattice QCD}
\author{Sinya Aoki
\thanks{\emph{Address after April 1, 2013:} Yukawa Institute for Theoretical Physics, Kyoto University, Kitashirakawa Oiwakecho, Sakyo-ku, Kyoto 606-8502, Japan}%
}                     
%
%
\institute{Graduate School of Pure and Applied Sciences, University of Tsukuba,  Ten-oh-dai 1-1-1, 
Tsukuba, Ibaraki 305-8571, Japan}
\date{Received: date / Revised version: date}
%
\abstract{
We review the potential method in lattice QCD, which has recently been proposed to extract  nucleon-nucleon interactions via numerical simulations.
We focus on the methodology of this approach by emphasizing the strategy of the potential method, the theoretical foundation behind it, and special numerical techniques. We compare the potential method with the standard finite volume method in lattice QCD, in order to make pros and cons of the approach clear. 
We also present several numerical results for the nucleon-nucleon potentials.
} 
\maketitle
\section{Introduction}
\label{intro}
Thank to both a steady growth of computational powers and various innovations of numerical algorithms,
we are now able to calculate static properties of light hadrons such as masses and decay constants at the physical point in the continuum limit of lattice QCD. See, for example,  Ref.~\cite{Fodor:2012gf} for a recent review.
One of the  next targets in lattice QCD calculations is to extract hadronic interactions such as the scattering between stable hadrons, masses and widths of unstable particles and binding energies of multi-hadron  states.
A standard framework to evaluate scattering phase shifts in lattice QCD is the L\"ushcer finite volume method\cite{Luscher:1990ux}, which relates the energy spectrum for two hadrons in a finite box with the elastic-scattering phase shift of two hadrons in the infinite volume. The method has been applied to various two-hadrons systems\cite{Orginos:2011zz}. 

An alternative but closely related approach to hadronic interactions in lattice QCD has recently been proposed
and applied to the nucleon-nucleon ($NN$) system\cite{Ishii:2006ec,Aoki:2008hh,Aoki:2009ji}. In the method,  
one first calculate the $NN$ potential, and  then extract
 physical observables such as the scattering phase shift by solving the Schr\"odinger equation with the potential obtained.
The method has been  widely applied to general hadronic interactions for 
baryon-baryon\cite{Nemura:2008sp,Nemura:2009kc,Inoue:2010hs,Inoue:2010es,Inoue:2011ai}, meson-baryon\cite{Ikeda:2010sg,Ikeda:2011qm,Kawanai:2010ev} and three nucleon\cite{Doi:2010yh,Doi:2011gq} systems,  mainly by the HAL QCD(Hadron to Atomic nuclei from Lattice QCD) collaboration.  See Refs.~\cite{Aoki:2011ep,Aoki:2012tk} for reviews on recent activities.

In this paper, we review this potential method, called the HAL QCD method from the name of the collaboration, focusing on the methodology of the approach such as the theoretical foundation and the numerical techniques.

\section{Strategy: HAL QCD method}
\label{sec:method}
In this section we explain the strategy of the HAL QCD method, taking the $NN$ system as an explicit example.

\subsection{Lippmann-Schwinger equation}
A concept of the potential in quantum field theories may appear in the Lippmann-Scwhinger equation\cite{text_weinberg}, 
\beqa
\vert \alpha \rangle_{\rm in} &=& \vert \alpha \rangle_0 + \int d\beta\, \frac{\vert \beta \rangle_0 T_{\beta\alpha}}{E_\alpha - E_\beta + i\epsilon},
\label{eq:LSE}
\eeqa
where the asymptotic in-state $\vert \alpha\rangle_{\rm in}$ satisfies
\beqa
(H_0 + V) \vert \alpha\rangle_{\rm in} &=& E_\alpha \vert\alpha\rangle _{\rm in},
\eeqa
while the non-interacting state $\vert \alpha\rangle_0$ does
\beqa
H_0 \vert \alpha\rangle_0 &=& E_\alpha \vert \alpha \rangle_0 .
\eeqa
 The off-shell $T$-matrix element $T_{\beta\alpha}$ is defined through the "potential" $V$ as
 \beqa
 T_{\beta\alpha} &=& {}_0\langle \beta \vert V \vert \alpha \rangle_{\rm in} .
 \eeqa
 This quantity is related to the on-shell $S$-matrix, $S=1- i T$, as
 \beqa
 {}_0\langle \beta \vert T \vert \alpha \rangle_0 &=& 2\pi \delta(E_\alpha-E_\beta)\, T_{\beta\alpha}.
 \eeqa
So our task is to extract $T_{\beta\alpha}$ in lattice QCD simulations. 
 
\subsection{Nambu-Bethe-Salpeter wave functions}
\label{eq:NBS}
The basic quantity in the HAL QCD method is the equal-time Nambu-Bethe-Salpeter(NBS) wave function\cite{Balog:2001wv}, defined for the $NN$ system as
\begin{eqnarray}
\Psi_{\alpha\beta,fg}^{\bks,s_1s_2}(\bx) &=& \langle 0 \vert T\left\{ N_{\alpha,f}(\br,0) N_{\beta,g}(\br+\bx,0) \right\} 
\nn \\ &\times & 
\vert NN, \bk, s_1s_2\rangle_{\rm in} ,
\label{eq:NBS}
\end{eqnarray}
where $\langle 0 \vert $ is the QCD vacuum state,  $T$ represents the time-ordered product, 
$\vert NN,\bk,s_1s_2\rangle_{\rm in}$ is the two-nucleon in-state which has helicity $s_1, s_2$ , the relative momentum $\bk$ and the total energy $W_{\bks}=2\sqrt{\bk^2+m_N^2}$ with the nucleon mass $m_N$ in the center of mass system.  For the interpolating operator for nucleon, we take 
the local one  given by $N_{\alpha,f}(x) =\epsilon_{abc} (u^a(x)^T C\gamma_5 d^b (x) ) q_{\alpha.f}^c(x)$ with $x=(\bx,t)$, where $a,b,c$ are color indices, $\alpha, f$ are spinor and flavor indices, $C=\gamma_2\gamma_4$ is the charge conjugation matrix, $q(x) =(u(x),d(x))$, and $u,d$ represent up and down quark fields, respectively.
As we will see, a potential in the HAL QCD scheme is defined through the NBS wave function, so that it depends on the choice of $N_{\alpha,f}(x)$. 

It is important to note that, as long as the total energy $W_{\bks}$ is below the pion production threshold such that
$W_{\bks } < W_{\rm th} \equiv 2m_N + m_\pi$ with the pion mass $m_\pi$,  
the NBS wave function at large $r\equiv \vert \bx \vert$ satisfies the Helmholtz equation as
\beqa
\left[k^2 +\nabla^2\right] \Psi_{\Gamma}^A(\bx) \simeq 0, \quad k=\vert \bk \vert,
\eeqa 
where we write $\Gamma=\alpha\beta,fg$ and $A=\bk,s_1s_2$ for simplicity. 
Furthermore, the radial part of the NBS wave function for a given orbital angular momentum $L$, the total spin $S$ and the total isospin $I$ for the  large $r$ is given by\cite{Ishizuka2009a,Aoki:2009ji}
\beqa
\Psi^{A}(r;LSI) &\propto&\frac{\sin(kr-L\pi/2+\delta_{LSI}(k))}{kr}e^{i\delta_{LSI}(k)} ,
\label{eq:NBS_asymp}
\eeqa
where $\delta_{LSI}(k)$ is the $NN$ scattering phase shift below the inelastic threshold, which appears in the $S$-matrix by the unitarity constraint. In the appendix, we derive
eq.~(\ref{eq:NBS_asymp}) for the scalar field, for simplicity,  using the Lippmann-Schwinger equation.

\subsection{Non-local potential}
Our task now becomes to extract the scattering phase shift encoded in the NBS wave function.
For this purpose, we define a non-local potential from the NBS wave function through the equation\cite{Ishii:2006ec,Aoki:2008hh,Aoki:2009ji}
\beqa
\left[E_k - H_0\right]\Psi_\Gamma^A(\bx) &=&\sum_{\Gamma_a}\int d^3 y\, U_{\Gamma,\Gamma_a}(\bx,\by)\Psi_{\Gamma_a}^A(\by)
\label{eq:potential}
\eeqa
where $E_k = k^2/(2\mu)$ with the reduced mass $\mu=m_N/2$, and $H_0=-\nabla^2/(2\mu)$.
First of all, the non-local potential $U(\bx,\by)$ is expected to be finite-ranged since massless particle exchanges between two nucleons are absent.
Secondly, the potential is finite and renormalization scheme independent, since the NBS wave function $\Psi_\Gamma^A(\bx)$ is multiplicatively renormalized in QCD and the same renormalization factor appears in both sides of eq.~(\ref{eq:potential}).
Thirdly, while Lorentz covariance is lost by taking the equal-time to define the NBS wave function in  eq.~(\ref{eq:NBS})   and the potential is defined through the non-relativistic Schr\"odinger equation ({\ref{eq:potential}), non-relativistic "approximation" has never been introduced to define $U(\bx,\by)$.

One of the most important properties for $U(\bx,\by)$ is that $U(\bx,\by)$ does not depend on energy (or more precisely momentum $\bk$) or helicities $s_1, s_2$ of the particular NBS wave function.
This can be shown by directly constructing such a non-local potential $U(\bx,\by)$ as\cite{Aoki:2009ji}
\beqa 
U_{\Gamma_a,\Gamma_b}(\bx,\by) &=& \sum_{A_a,A_b}^{\vert \bks_{a,b}\vert  < k_{\rm th}}
\left[E_{k_a}- H_0\right] \Psi_{\Gamma_a}^{A_a}(\bx)
 {\cal N}^{-1}_{A_a,A_b} \Psi_{\Gamma_b}^{A_b}(\by),\nn\\
 \label{eq:nonLpotential}
\eeqa
where $A_{a} = \bk_a, s^a_1s^a_2 $ etc. and $k_{\rm th}$ is the threshold momentum which satisfies
$W_{\rm th} = 2\sqrt{k_{\rm th}^2+m_N^2}$, so that the summations over $\bk_a$ and $\bk_b$ are restricted below the inelastic threshold. Here ${\cal N}^{-1}$ is the inverse of ${\cal N}$ defined from the inner product of the NBS wave functions as
\beqa
{\cal N}^{A_a,A_b}  &=& \left(\Psi^{A_a},\Psi^{A_b}\right) \equiv \sum_{\Gamma}
\int d^3x\, \Psi_{\Gamma}^{A_a}(\bx)^\dagger \Psi_{\Gamma}^{A_b}(\bx), \nn
\eeqa
and therefore the inverse satisfies
\beqa
\sum_{A_c}^{\vert \bks_c\vert < k_{\rm th}} {\cal N}_{A_b, A_c}^{-1} {\cal N}^{A_c, A_b} 
&=& \delta_{A_a}^{A_b} \equiv \delta^{(3)}(\bk_a-\bk_b)\delta_{s^a_1}^{s^b_1}\delta_{s^a_2}^{s^b_2}, \nn
\eeqa
for $\vert \bk_{a,b}\vert < k_{\rm th}$.  It is easy to see that this $U(\bx,\by)$ is energy($\bk$) independent by construction and satisfies eq.~(\ref{eq:potential}) as
\beqa
&&\sum_{\Gamma_b} \int d^3y\, U_{\Gamma,\Gamma_a}(\bx,\by) \Psi_{\Gamma_a}^{A}(\by)
= \sum_{A_a,A_b}^{\vert \bks_{a,b}\vert  < k_{\rm th}}
\left[E_{k_a}- H_0\right] \nn \\
&\times& \Psi_{\Gamma}^{A_a}(\bx) \ {\cal N}^{-1}_{A_a,A_b} {\cal N}^{A_b,A} = \left[E_k-H_0\right]\Psi_{\Gamma}^A(\bx) 
\eeqa
for $\vert \bk\vert < k_{\rm th}$. 
Once the non-local potential $U(\bx,\by)$ which satisfies eq.~(\ref{eq:potential}) is obtained,
we can extract the phase shift $\delta_{LSI}(k)$ at $k< k_{\rm th}$  in QCD
under the property in eq.~(\ref{eq:NBS_asymp}),  by solving the Schr\"odinger equation with this potential. 
Since the potential itself is not a physical observable, however, the potential which satisfies eq.~(\ref{eq:potential}) is not unique. One may add terms which affect eq.~(\ref{eq:potential}) only
above the inelastic threshold ($k > k_{\rm th}$) while keeping eq.~(\ref{eq:potential}) intact below the inelastic threshold.

\subsection{Velocity expansion}
The construction of $U(\bx,\by)$ given in eq.~(\ref{eq:nonLpotential}) is important to prove that such a energy-independent potential indeed exists. In lattice QCD simulations, however, the NBS wave functions can be obtained only for a ground state and possibly a few low-lying excited states, so that a summation over $A_a$ becomes inaccurate.  For practical uses, we therefore expand the non-local potential in terms of velocity (or derivative) and determine local coefficient functions order by order of the expansion\cite{Ishii:2006ec,Aoki:2008hh,Aoki:2009ji}.
Explicitly we have $U(\bx,\by) = V(\bx,\nabla)\delta^{(3)}(\bx-\by)$ with
 \beqa
 V(\bx,\nabla) &=& \underbrace{V_0(r) + V_\sigma(r)\bsigma_1\cdot \bsigma_2+V_T(r) S_{12}}_{=V^{\rm LO}(\bx)}\nn \\
  &+& \underbrace{V_{\rm LS}(r) \bL\cdot \bS}_{=V^{\rm NLO}(\bx)}  + O(\nabla^2),\quad
 r=\vert \bx \vert,
 \label{eq:V_expansion}
 \eeqa
where $\bsigma_i$ is the Pauli-matrix acting on the spin index of the $i$-th nucleon, $S_{12} = 3(\bx\cdot\bsigma_1)(\bx\cdot\bsigma_2)/r^2 -\bsigma_1\cdot\bsigma_2$ is the tensor operator, $\bL= \bx\times\bp$ is an orbital angular momentum, and $\bS=(\bsigma_1+\bsigma_2)/2$ is the total spin. 
Each local coefficient function is further decomposed as $ V_X(r)=V_X^0(r) + V_X^\tau(r)\, \btau_1\cdot\btau_2$ for $X=0,\sigma,T,{\rm LS},\cdots$, where $\btau_i$ is the flavor matrix acting on the flavor index of the $i$-th nucleon.
This form of the velocity expansion agrees with the result obtained by the symmetry\cite{okubo1958}.

We can determine these coefficient functions from NBS wave functions with particular quantum numbers, order by order in the velocity expansion.
Once the potential is obtained at some order of the expansion, we can solve the Schr\"odinger equation with this potential in order to extract physical observables such as the scattering phase shift and the binding energy of the deuteron.

\subsection{Remarks}
Let us give some remarks on the HAL QCD method.

First of all, potentials themselves are not physical observables, and they are therefore not unique. For example, potentials depend on the choice of nucleon operators to define the NBS wave functions. We adopt here the local nucleon operator where all three quarks are put on the same point, since it is a convenient choice for the reduction formula of composite particles\cite{Nishijima,Zimmermann,Haag}, but  other choices are equally possible.
We consider such a dependence of a potential on its definition  as a "scheme" of the potential.
While the potential is therefore scheme dependent, physical observables such as the scattering phase shift and the binding energy are of course physical thus scheme independent.

Although potentials are scheme dependent, they are useful to understand or describe "physics" of hadronic interactions. For example, the $NN$ potential best  summarizes the $NN$ scattering at higher energy in terms of its repulsive core,  as the running coupling constant in QCD, which is also scheme dependent,  describes the deep inelastic scattering data at high energy well in terms of its asymptotic freedom.

Among different schemes (of potentials or running couplings constant), some schemes are better than others.
While a good convergence of the perturbative expansion for a certain class of observables is a reasonable criteria for a good  running coupling constant,  a good convergence of the velocity expansion, which means weak non-locality, is a reasonable criteria for a good potential. In this sense, a completely local and energy-independent potential  would be the best one  if no inelastic threshold were present.

It is also important to note that the convergence of the velocity expansion of the potential can be examined within the HAL QCD method. For example, if we have $\Psi^{A_n}$  for $n=1,2,\cdots, N$, from which we can determine $N-1$ local functions in the velocity expansion in $N$ different ways.
A variation among $N$ different ways  gives an estimate of  the size of higher order  terms neglected.
Furthermore we can determine one of these higher order terms using all $N$ NBS wave functions.
A convergence of the velocity expansion will be  considered later. 

\section{Lattice methods}
In this section, we consider an explicit procedure to extract potentials in lattice QCD simulations.

\subsection{An extraction of NBS wave functions from correlation functions}
We first consider a method to extract NBS wave functions from 4-pt  correlation functions, defined by
\beqa
F_\Gamma (\bx, t-t_0) &=& \langle 0 \vert T\{N_{\alpha,f}(\br,t) N_{\beta,g}(\br+\bx,t)\} \overline{\cal J}(t_0) \vert 0 \rangle
\nn
\eeqa  
for $t > t_0$, where $\Gamma=\alpha\beta,fg$, and 
$\overline{\cal J}(t_0)$ is a source operator which creates two-nucleon states and its explicit form will be given later.
By inserting a complete set in the above definition, we obtain
\beqa
&&F_\Gamma (\bx, t-t_0) = \langle 0 \vert T\{ N_{\alpha,f}(\br,t) N_{\beta,g}(\br+\bx,t)\} \nn \\
& & \sum_{n,s_1s_2} \vert NN, A_n \rangle_{\rm in} \times {}_{\rm in}\langle NN, A_n \vert  \overline{\cal J}(t_0) \vert 0 \rangle + \cdots 
\nn \\
&=& \sum_{n,s_1s_2} Z(A_n) \Psi^{A_n}_{\Gamma}(\bx) e^{-W_{\bks_n} (t-t_0)} +\cdots ,
\eeqa
where $A_n=\bk_n,s_1s_2$, 
$
Z(A_n) = {}_{\rm in}\langle NN, A_n \vert  \overline{\cal J}(0) \vert 0 \rangle
$,
and ellipses represent inelastic contributions.

As in the standard method to extract hadron masses in lattice QCD, we extract the NBS wave function for the ground state from the above correlation function by taking $(t-t_0)\rightarrow \infty$ as
\beqa
F_\Gamma (\bx, t-t_0) &\simeq& Z(A_0)\Psi^{A_0}_{\Gamma}(\bx) e^{-W_{\bks_0} (t-t_0)},
\label{eq:GS}
\eeqa
where $W_{\bks_0}$ is the lowest energy of $NN$ states.

The extraction of the NBS wave function in eq.~(\ref{eq:GS}) relies on the ground state saturation for the correlation function, which can in principle be achieved by taking a large $t-t_0$. In practice, however, 
it is difficult to realize the ground state saturation for the $NN$ system within reasonable errors,
since $F_\Gamma (\bx, t-t_0)$ becomes very noisy at large $t-t_0$.

\subsection{An improved extraction}
The signal-to-noise ratio for the nucleon 4-pt function $F_\Gamma$ behaves for large $t$ as
$
{\cal S}/{\cal N} \sim e^{-2(m_N-3m_\pi/2)t}
$ \cite{Parisi:1983ae,Lepage:1989hd},
which decreases for  lighter pion masses toward its physical value. 
Furthermore,  as we increase the volume, the splitting between the ground state and the 1st excited state for the $NN$ system becomes smaller as
$
\Delta E \simeq \bk_{\rm min.}^2/{m_N} = \left({2\pi}/{L}\right)^2/m_N,
$
which requires lager $t$ for the ground state saturation.
The behavior of statistical noise in the above, however, makes the signals very poor at such large $t$ for the $NN$ system.

An improved extraction of the NBS wave function has  recently been proposed to overcome the above difficulties\cite{HALQCD:2012aa}.
We first normalize the 4-point correlation function as
\begin{eqnarray}
R_\Gamma (\bx,t) &\equiv&\frac{ F_\Gamma (\bx,t)}{ (e^{-m_N t} )^2} 
\simeq \sum_{n,s_1,s_2} Z(A_n)\  \Psi^{A_n}_\Gamma(\bx)e^{-\Delta W_{\bks_n} t} \nn
\end{eqnarray}
where $\Delta W_{\bks} = 2\sqrt{m_N^2+ \bk^2} - 2 m_N$. 
Using an identity $\Delta W_{\bks} = \bk^2/m_N - (\Delta W_{\bks})^2/(4m_N)$ and neglecting inelastic contributions, 
we obtain the time-dependent Schr\"odinger-like equation
\begin{eqnarray}
&&\left\{-H_0 - \frac{\partial}{\partial t} +\frac{1}{4m_N}\frac{\partial^2}{\partial t^2}\right\}R_\Gamma(\bx,t) \nn \\
&=& \sum_{\Gamma_a} \int d^3 y\,  U_{\Gamma\Gamma_a}(\bx,\by) R_{\Gamma_a} (\by,t) 
\label{eq:t-dep1} \\
&\simeq &  \sum_{\Gamma_a} V_{\Gamma\Gamma_a}(\bx) R_{\Gamma_a}(\bx,t) +\cdots ,
\label{eq:t-dep2}
\end{eqnarray}
which shows that the same $U(\bx,\by)$ in eq.~(\ref{eq:potential}) can be obtained  from $R_\Gamma(\bx,t)$.
An advantage of this method is that the ground state saturation is no more required for $R_\Gamma(\bx,t)$ to satisfy eq.~(\ref{eq:t-dep1}) or eq.~(\ref{eq:t-dep2}).
For this method to work, however, $t$ has to be large enough  that elastic contributions dominate $R_\Gamma(\bx,t)$. 

In the scattering of two different particles such as the $N\Xi$ scattering, 
we must employ the non-relativistic expansion as $ \Delta W_{\bks} =\sqrt{\bk^2+m_N^2}+\sqrt{\bk^2+m_\Xi^2}-m_N-m_\Xi \simeq \bk^2/(2\mu)$ with the reduced mass $\mu= m_N m_\Xi/(m_N+m_\Xi)$. Within this approximation, we obtain
\beqa
\left\{-H_0 - \frac{\partial}{\partial t} \right\}R^{N\Xi}_\Gamma(\bx,t) &\simeq&
\sum_{\Gamma_a}V_{\Gamma\Gamma_a}(\bx)  R^{N\Xi}_{\Gamma_a} (\by,t) . 
\eeqa

\subsection{Source operators}
We choose the source operator ${\cal J}$ so as to fix quantum numbers of the state $\vert NN, A_n\rangle$.
Instead of the $SO(3,\bR)$,  
we classify the states on the hyper cubic lattice in terms of the irreducible representation of the cubic transformation $SO(3,\bZ)$, denoted by $A_1,A_2,E,T_1,T_2$, whose dimensions are 1, 1, 2, 3, 3. 
In Table~\ref{tab:decomp}, the orbital angular momentum $L$ representation in $SO(3,\bR)$ is decomposed in terms of irreducible representations in $SO(3,\bZ)$ at $L\le 6$.
For example, the table says that  the source operator ${\cal J}(t_0)$ in the $A_1$ representation with positive parity produces states with $L=0,4,6,\cdots$ at $t=t_0$ from the vacuum, while that in the $T_1$ with negative parity  produces states with $L=1,3,5,\cdots$ .  
\begin{table}
\caption{A decomposition of irreducible representations of $SO(3,\bR)$ with the orbital angular momentum $L$ in terms of $SO(3,\bZ)$ representations. Here $P=(-1)^L$ represents parity.}
\label{tab:decomp}       
\begin{center}
\begin{tabular}{lllllll}
\hline\noalign{\smallskip}
$L$ & $P$ & $A_1$ & $A_2$ & $E$ & $T_1$ & $T_2$   \\
\noalign{\smallskip}\hline\noalign{\smallskip}
$0(S)$ & $+$ & 1 & 0 & 0 & 0 & 0 \\
$1(P)$ & $-$ & 0 & 0 & 0 & 1 & 0 \\
$2(D)$ & $+$ & 0 & 0 & 1 & 0 & 1 \\
$3(F)$ & $-$ & 0 & 1& 0 & 1 & 1 \\
$4(G)$ & $+$ & 1 & 0 & 1 & 1 & 1 \\
$5(H)$ & $-$ & 0 & 0 & 1 & 2 & 1 \\
$6(I)$ & $+$ & 1 & 1 & 1 & 1 & 2 \\
\noalign{\smallskip}\hline
\end{tabular}
\end{center}
\end{table}
The total spin $S$ for two nucleons becomes $1/2\otimes 1/2 = 1\oplus 0$, corresponding to $T_1(S=1)$ and $A_1(S=0)$ of the $SO(3,\bZ)$, respectively.
Thus the total "angular momentum" $J$ for a two nucleon system is given by the product of $R_1\otimes R_2$, where $R_1 =A_1,A_2,E,T_1,T_2$ for the orbital "angular momentum" while $R_2=A_1,T_1$ for the total spin. 
In Table~\ref{tab:product}, a decomposition of the product $R_1\otimes R_2$ is given in terms of the direct sum.
\begin{table}
\caption{A decomposition of the direct product $R_1\otimes R_2$ in terms of the direct sum of irreducible representations. By definition $R_1\otimes R_2 = R_2\otimes R_1$.
}
\label{tab:product}       
\begin{center}
\begin{tabular}{ll | c || l l | c l}
\hline
$R_1$ & $R_2$ & $R_1\otimes R_2$ &  $R_1$ & $R_2$ & $R_1\otimes R_2$  \\
\hline
$A_1$ & $A_1$ & $A_1$ &$E$ & $E$ & $A_1\oplus A_2\oplus E$  \\
$A_1$ & $A_2$ & $A_2$  &  $E$ & $T_1$ &  $T_1\oplus T_2$\\
$A_1$ & $E$ & $E$  & $E$ & $T_2$ & $T_1\oplus T_2$\\
$A_1$ &$T_1$ & $T_1$  &  $T_1$ & $T_1$ &  $A_1\oplus E \oplus T_1\oplus T_2$\\
$A_1$ & $T_2$ & $T_2$ &$T_1$ & $T_2$ &  $A_2\oplus E \oplus T_1\oplus T_2$  \\
$A_2$ & $A_2$ & $A_1$ &$T_2$ & $T_2$ &  $A_1\oplus E \oplus T_1\oplus T_2$\\
$A_2$ & $E$ & $E$ &&&\\
$A_2$ & $T_1$ & $T_2$&&& \\
$A_2$ & $T_2$ & $T_1$ &&&\\
\hline
\end{tabular}
\end{center}
\end{table}

Most of the HAL QCD results are obtained by the wall source defined by
\beqa
{\cal J}_{\alpha\beta,fg}^{\rm wall} (t) = N_{\alpha,f}^{\rm wall}(t) N_{\beta,g}^{\rm wall}(t), 
\eeqa
where $\alpha,\beta =1,2 $ are (upper) spinor indices, $f,g$ are flavor indices, and $N^{\rm wall}(t)$ is obtained by replacing local quark field $q(x)$ of $N(x)$ with the wall quark field, $q^{\rm wall}(t) \equiv \sum_{\bxs} q(\bx,t)$ together with the Coulomb gauge fixing only at the time slice of the source.

The source operator $\overline{\cal J}^{\rm wall}(t_0)$ creates states with the zero angular momentum at $t=t_0$,
which belongs to the $A_1$ representation with positive parity $P=+$. 
Using the spin projection $P^{(S)}$ with $P^{S=0}=\sigma_2, P^{S=1}=\sigma_1 $,
we fix the total angular momentum $J$ with $J_z=0$ and the total isospin $I$ of the source as
\beqa
{\cal J}(t_0; J^{P=+},I) &=& P_{\beta\alpha}^{(S)} {\cal J}_{\alpha\beta,fg}(t_0), 
\eeqa
where the total isospin $I$ is automatically fixed once the total spin is given due to the fermonic nature of nucleons: $(S,I) =(0,1)$ or $(1,0)$.
Since $A_1^+(L=0) \otimes A_1 (S=0) = A_1^+$ or $A_1^+(L=0) \otimes T_1 (S=1) = T_1^+$, 
the state with $(J^P,I) =(A_1^+,1)$ for the spin-singlet or $(J^P,I) =(T_1^+,0)$ for the spin-triplet  is created at $t=t_0$ by the above source. As these quantum numbers are conserved by QCD interactions, the NBS wave function extracted at $t > t_0$ has the same $(J^P,I)$. Moreover,  as a speciality of the two nucleon system with equal up and down quark masses, the total spin $S$ is also conserved at $t> t_0$:
The constraint $(-1)^{(S+1)+(I+1)} P = -1$ should be satisfied due to the fermionic nature of nucleons, while parity $P$ and the isospin $I$ are conserved in QCD with equal up and down quark masses. Therefore $S$ is conserved under the condition that $S=0,1$.
The orbital angular momentum $L$, however, is not conserved in general. 
While the state with $(J^P,I)=(A_1^+,1)$ has $L=A_1^+$ even at $t > t_0$, the other state with $(J^P,I)=(T_1^+,0)$ has $L=A_1^+$ and $L=E^+\oplus T_2^+$ components at $ t> t_0$, corresponding  to $L=0$ (S-wave) and $L=2$ (D-wave) in $SO(3,\bR)$. 
Note that Table~\ref{tab:product} tells us that not only $L=A_1^+, E^+\oplus T_2^+$ components but also the $L=T_1^+$ component exist in the state with $(J^P,I)=(T_1^+,0)$. The latter extra component is expected to be small since it appears as a consequence of the violation of rotational symmetry on the cubic lattice.

The orbital angular momentum $L$ of the NBS wave function can be projected onto a particular value by the operator $P^{(L)}$  as
\beqa
\Psi^A (\bx; J^P,I,L) &=& P^{(L)}  \Psi^A(\bx; J^P,I)
\eeqa
where the total spin is given by $S=1-I$, and $\Psi^A(\bx; J^P,I)$ is extracted from the 4-pt function generated by the source ${\cal J}(t_0;J^P,I)$ as
\beqa
F_\Gamma(\bx,t-t_0;J^P,I) &\simeq & Z(A,J^P,I) \Psi^A_\Gamma (\bx;J^P,I) e^{- W_{\bks}(t-t_0)}, \nn \\
Z(A, J^P,I) &=& {}_{\rm in}\langle NN, A \vert \overline{\cal J}(0;J^P,I)\vert 0\rangle
 \eeqa
 for large $t-t_0$.  The projection $P^{(L)}$ is defined for an arbitrary function $\varphi(\bx)$ by
 \beqa
 P^{(L)}\varphi(\bx) &=& \frac{d_L}{24} \sum_{g\in SO(3,\bZs)} \chi^L(g) \varphi(g^{-1}\bx) 
 \eeqa
 for $L=A_1,A_2,E,T_1,T_2$, where $\chi^L$ denotes the character for $L$, $g$ is one of 24 elements of $SO(3,\bZ)$ and $d_L$ is the dimensions of $L$. It is also noted that  the $A_1$ state may contain $L=4,6,\cdots$ components other than the dominant $L=0$ component.
 
\subsection{Local potentials at the leading order}
Local potentials at the leading order (LO) of the velocity expansion take the form 
\beqa
V^{\rm LO}(r) &=& V_0(r) + V_\sigma(r) \bsigma_1\cdot \bsigma_2 + V_T(r) S_{12}.
\eeqa

For the isospin-triplet (spin-singlet) sector, $S_{12}=0$ and $\bsigma_1\cdot \bsigma_2 = -3$ imply
\beqa
V_C^{I=1}(r) 
&=& \frac{\left[E_k-H_0\right]\Psi^A(\bx; A_1^+,1)}{\Psi^A(\bx; A_1^+,1)},
\eeqa
where $ V_C^{I=1}(r)=V_0^{I=1}(r) - 3 V_\sigma^{I=1}(r)$, which is often referred to as the central potential for the ${}^1S_0$ state, where  the notation $^{2S+1}L_J$ is used. It is noted, however, that the potentials at the LO of the velocity expansion do not depend on the quantum numbers of the state, $J$ and $L$.
In this sense, it is more precise to say that $V_C^{I=1}(r)$ is the parity-even isospin-triplet (spin-singlet) potential determined from the state with $J=L=A_1$. Determinations of this potential from other states can give estimates for the size of contributions from higher order terms in the velocity expansion.

For the isospin-singlet (spin-triplet) sector, both tensor potential $V_T$ and central potential $V_C$ appear at the LO.
The Schr\"odinger equation for the state with $(J^P,I)=(T_1^+,0)$ reads
\beqa
&&\left[H_0 +V_C^{I=0}(r) + V_T^{I=0}(r) S_{12}\right] \Psi^A(\bx; T_1^+,0)  \nn \\
&=& 
E_k \Psi^A(\bx; T_1^+,0),
\eeqa
where $V_C^{I=0}(r) =V_0^{I=0}(r)+V_\sigma^{I=0}(r)$. 
With projections onto $A_1$ and $E\oplus T_2$ components, we have $V_C^{I=0}$ and $V_T^{I=0}$ as
\beqa
V_C^{I=0}(r)&=& E_k -\frac{1}{\Delta(\bx)}\left(  [{\cal Q} S_{12} \Psi^A] (\bx)  H_0 [{\cal P}  \Psi^A] (\bx) \right.  \nn \\
&-&\left. [{\cal P} S_{12} \Psi^A](\bx)  H_0 [{\cal Q}  \Psi^A](\bx) \right) , 
\label{eq:V3S1}\\
V_T^{I=0}(r) &=& \frac{1}{\Delta(\bx)}\left(  [{\cal Q}  \Psi^A](\bx)  H_0 [{\cal P}  \Psi^A](\bx) \right.  \nn \\&-&
 \left. [{\cal P}  \Psi^A](\bx)  H_0 [{\cal Q}  \Psi^A](\bx) \right),
 \label{eq:VT}
 \eeqa
\beqa
\Delta(\bx) \equiv [{\cal Q} S_{12} \Psi^A](\bx)  [{\cal P}  \Psi^A](\bx) 
- [{\cal P} S_{12} \Psi^A](\bx)  [{\cal Q}  \Psi^A](\bx), \nn
\eeqa 
where
${\cal P}\Psi^A(\bx) =P^{(A_1)} f(\bx)$ and  ${\cal Q}\Psi^A(\bx) =P^{(E\oplus T_2)} f(\bx)$ with
$f(\bx) \equiv \Psi^A_\Gamma(\bx; T_1^+,0)$.
In numerical simulations, $\Gamma =(\alpha\beta,fg) =(2,1,2,1)$ is mainly employed, and the approximation that ${\cal Q} \simeq 1-{\cal P}$ is used by neglecting small $T_1$ component.
We may define the effective central potential as
\beqa
V_{C,{\rm eff}}^{ I=0} (r) &=& \frac{[E_k-H_0]{\cal P}\Psi^A(\bx)}{{\cal P}\Psi^A(\bx)},
\eeqa
which differs from $V_C^{I=0}(r)$ by $O(V_T^2)$ in the second order perturbation for small $V_T$.

\subsection{A comparison }
We here briefly compare the potential method with the direct extraction of the phase shift via the L\"usher's finite volume method in lattice QCD.

The potential method by construction gives the correct phase shift at  a particular $k$ where the NBS wave function is calculated, while the phase shift at other values of $k$ is approximated one due to the velocity expansion of the non-local potential.
With this systematic uncertainty in mind, the potential method can reveal the global structure of the phase shift in the wide range of (continuous) $k$, while the finite volume method can gives the exact phase shift at a few discrete points of $k$.

The finite size correction to the potential is expected to be small, since no massless particle exchange exists between nucleons. Indeed, the L\"usher's finite volume method assumes that the potential remains intact  as long as the volume is large enough so that the interaction range of the potential is smaller then the half of the lattice extension, $L/2$. Under this condition, the scattering wave satisfies the free 
Schr\"odinger equation outside the interaction range. Due to the (periodic) boundary condition, we have the discrete values of $k$ in the finite box,  which give some information of hadronic interactions\cite{Luscher:1990ux}. 
To extract the phase shift at the corresponding values of $k$, we need an additional assumption:
Let us consider the allowed values of $k$ for the state with  $L=A_1$, which contains not only $L=0$ components but also $L=4,6,\cdots$, contributions. We then extract the phase shift $\delta_L(k)$ for the $L=0$ partial wave, under the assumption that the $L=0$ component dominates in this state.
In the case of the potential method, on the other hand, we do not need such an assumption. Once the potential is obtained, we can calculate the phase shift for an arbitrary $L$ by solving the  Schr\"odinger
equation in the {\it infinite} volume, again with the systematic uncertainty  of the velocity expansion.

We also expect that the quark mass dependence of the potential is much milder than that of physical observables such as the scattering length. While the $NN$ scattering length is small in the heavy quark mass region, it diverges when the deuteron bound state is formed at a lighter quark mass\cite{Kuramashi:1995sc}.
Therefore the scattering length varies from almost zero to infinity as the quark  mass decreases.
Such a drastic change of the scattering length can be realized by a small change of the potential shape as a function of the quark mass.  This would make chiral extrapolations for the potentials to the physical pion mass more stable than those for the scattering length.

\section{Lattice results}
In this section, numerical results for nuclear potentials obtained by the HAL QCD collaboration are introduced.

\subsection{Nuclear potentials in full QCD by the improved extraction}
We evaluate nuclear potentials,  employing  (2+1)-flavor  QCD  gauge
configurations  generated by PACS-CS  collaboration \cite{Aoki:2008sm}
on $32^3\times 64$  lattice with the RG improved  Iwasaki gauge action
at  $\beta = 1.9$  and the  non-perturbatively $O(a)$  improved Wilson
quark  action with $C_{\rm SW} =  1.715$, which corresponds  to the lattice
spacing  $a\simeq  0.091$  fm  ($a^{-1}=2.176(31)$ GeV),  the  spatial
extent $L =  32a \simeq 2.90$ fm.   We calculate $R(\bx,t)$ at a fixed value of light and strange quark mass combination, which corresponds to 
$m_{\pi} \simeq  701$ MeV, $m_K\simeq 789$ MeV and $m_{N}
\simeq 1583$ MeV. 

The periodic boundary  condition is used for spatial directions, while
the Dirichlet  boundary condition is taken for  the temporal direction
at  $t_{\rm DBC}=32  a$ and $-32 a$,
to avoid opposite propagations  of two nucleons in temporal direction,
i.e, one  propagates forward and the other  propagates backward.  From
time-reversal and  charge conjugation symmetries, we  can average over
forward propagation at $t>0$ and backward propagation at $t< 0$ with a
wall source at $t=0$.
By temporally shifting gauge configurations, 21 source points are used
per  one configuration and  390 gauge  configurations are  employed in
total.  Statistical errors are  estimated by the Jackknife method with
a bin size  of 10 configurations.   In our actual  calculation, we replace $e^{-m_{N}t}$ in the denominator of $R(\bx,t)$ by the single-nucleon propagator
$C_{N}(t) \equiv \sum_{\vec x}\langle 0\vert T[N(x)\bar{N}(0)]\vert 0\rangle$.
Time derivatives  are evaluated after
applying the polynomial interpolation of degree 5 to $R(\bx, t)$.

\begin{figure}
\resizebox{0.5\textwidth}{!}{%
  \includegraphics{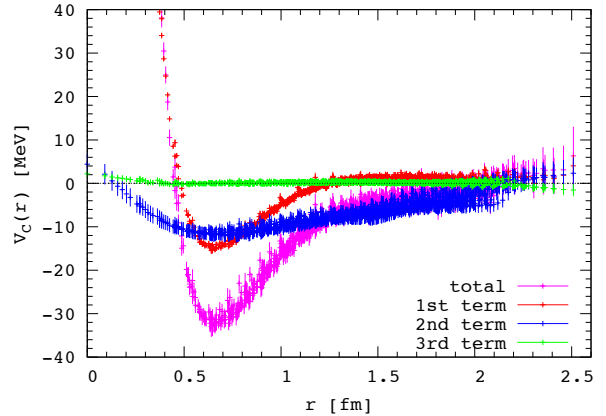}
}
\caption{Three contributions to $V_C^{I=1}(r)$ in eq.~(\ref{eq:breakup}) at $t=9$.
Taken from Ref.~\cite{HALQCD:2012aa}. }
\label{fig:singlet}       
\end{figure}
In the improved extraction, the central potential $V_C^{I=1}(r)$ at the LO is given by
\beqa
V_C^{I=0}(r) &=& -\frac{H_0 R(\bx,t)}{R(\bx,t;)} -\frac{(\partial/\partial t) R(\bx,t)}{R(\bx,t)} \nn \\
&+&\frac{1}{4m_N}\frac{
(\partial/\partial t)^2 R(\bx,t)}{R(\bx,t)},
\label{eq:breakup}
\eeqa
which are shown separately in Fig.~\ref{fig:singlet} at $t=9$\cite{HALQCD:2012aa}.
The first term in eq.~(\ref{eq:breakup}) (the red points) determines the overall shape of the potential, while the second term (the blue points) gives a major correction.
The  third term (the green points) corresponding to the relativistic correction,
on the other hand,  is
negligible, showing  that  the  non-relativistic
approximation  $\Delta W_{\bks}  \simeq   \bk^2/m_N$  works well.
Note that the second term in eq.~(\ref{eq:breakup}) would be constant if the ground state saturation were achieved. A clear $r$-dependence of the second term tells us that contaminations of excited states indeed exist and are non-negligible.

The potential in Fig.~\ref{fig:singlet} has a similar structure to the know phenomenological $NN$ potentials,
namely the repulsive core at short distance surrounded by the attractive well at medium and long distances,
as shown in Fig.~\ref{fig:phen-pot}. The first result for the NN potential, obtained in quenched QCD by the HAL QCD method\cite{Ishii:2006ec},  also reproduces this structure, and this success  has received general recognition\cite{nature}.
Note however that lattice artifacts may be large at very short distance ( {\it i.e.} $ r \le 0.1$ fm).
\begin{figure}
\resizebox{0.4\textwidth}{!}{%
  \includegraphics{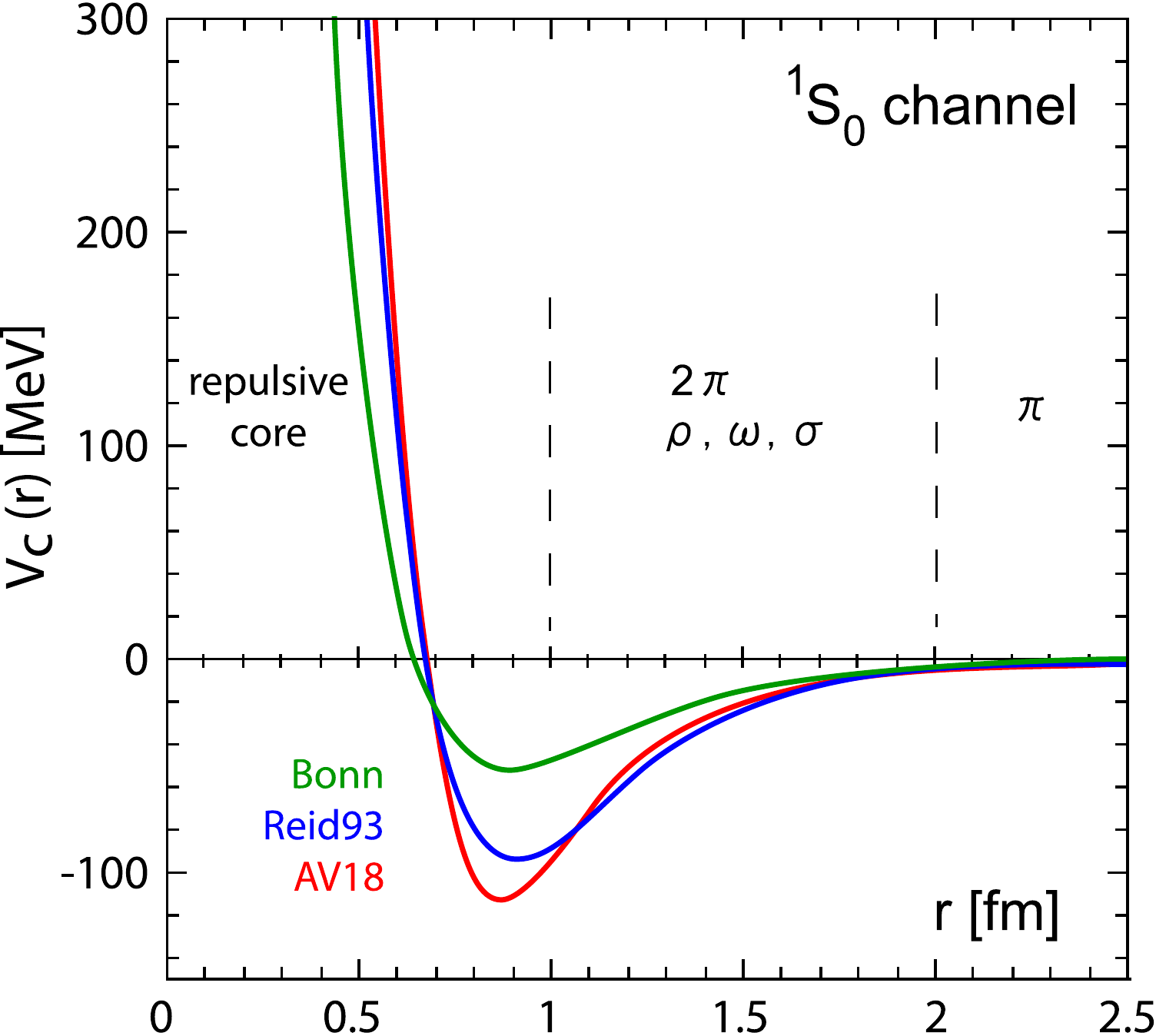}
}
\caption{Three examples of  the  phenomenological $NN$ potential in the isospin-triplet (spin-singlet) sector ($V^{I=1}_C(r)$ ),  Bonn\cite{Machleidt:2000ge}, Reid93\cite{Stoks:1994wp} and Argonne $v_{18}$\cite{Wiringa:1994wb}.  Taken from Ref.~\cite{Ishii:2006ec}.  }
\label{fig:phen-pot}       
\end{figure}

\subsection{Scattering phase shift}
\begin{figure}
\resizebox{0.5\textwidth}{!}{%
  \includegraphics{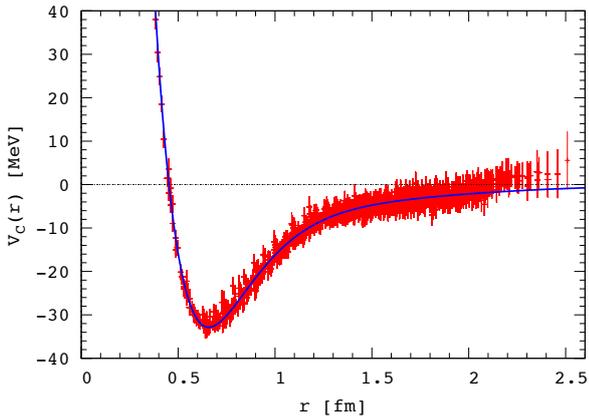}
}
\caption{The  multi-Gaussian fit  of  the  central
  potential $V^{I=1}_C(r)$ with $N_{\rm  Gauss} = 5$ at $t=9$. Taken from Ref.~\cite{HALQCD:2012aa}.  }
\label{fig:fit}       
\end{figure}

\begin{figure}
\resizebox{0.5\textwidth}{!}{%
  \includegraphics{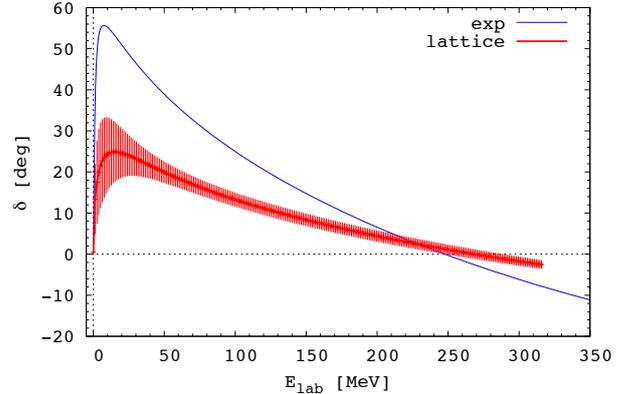}
}
\caption{The  scattering  phase in  $^1S_0$  channel in  the
  laboratory  frame obtained  from the  lattice $NN$  potential, together
  with experimental data\cite{nn-online}.  Taken from Ref.~\cite{HALQCD:2012aa}. }
\label{fig:scattering.phase}       
\end{figure}
To calculate the $NN$ scattering phase shift by solving the Schr\"odinger equation with the potential  in the infinite volume,
the central  potential $V^{I=1}_C(r)$  is fitted with multi-Gaussian functions as  $
  g(r) 
  \equiv
  \sum_{n=1}^{N_{\rm gauss}}
  V_n \cdot \exp(-\nu_n r^2),
$
where  $V_n$ and  $\nu_n  (>0)$  are used  as  fit parameters,  $N_{\rm
  gauss}$  denotes the number of Gaussian functions.
The fit  with multi-Gaussians but without a Yukawa-function works well, as  shown in Fig.~\ref{fig:fit}\cite{HALQCD:2012aa}, presumably due to the heavy pion mass.

We then solve the Schr\"odinger equation in  $^1S_0$ channel with this potential,
in order to extract  the  scattering   phase $\delta(k)$, which is shown in
Fig.~\ref{fig:scattering.phase}, together  with the experimental  data  for  comparison\cite{HALQCD:2012aa}.   
Qualitative feature of the phase shift  as a function of $k$ 
is well reproduced, though the strength is weaker, most likely due to 
the heavy pion mass ($m_\pi \simeq 701$ MeV) in this calculation.
In fact, the recent 3-flavor QCD simulations
 show that the  $NN$ phase shift approaches toward the physical value  
as the quark mass  decreases \cite{Inoue:2011ai}.
The  scattering  length at $m_\pi \simeq 701$ MeV in the present method, 
calculated  from the  derivative  of  the
scattering phase  shift at $E_{\rm lab} =  0$, leads  to
 $a(^1S_0) =  \lim_{k\to 0} \tan \delta(k)/k = 1.6\pm 1.1$ fm, which is still smaller than the experimental value at the physical point, $a^{\rm (exp)}(^1S_0)\sim 20$ fm ( strong attractive in our sign convention), as seen from a comparison in Fig.~\ref{fig:scattering.phase}.

\subsection{Tensor potential}
Using the same gauge configurations generated by the PACS-CS collaboration, the LO potentials for the isospin-singlet (spin-triplet)  have been extracted.  In Fig.~\ref{fig:full} we show $V_C^{I=0}(r)$ and $V_T^{I=0}(r)$, together with  $V_C^{I=1}(r)$ for a comparison\cite{Ishii:2010th}.
While central potentials for both sectors look  similar, the tensor potential $V_T^{I=0}(r)$ is negative  for the whole range of $r$, so that no repulsive core appears in this sector.
The tensor potential seems finite at $r=0$,  but we have to be careful to conclude such a short distance behavior  of the potential, since lattice artifacts are large at short distances. 

The meson theory predicts that  the tensor potential receives a significant contribution from one-pion exchange, so that $V_T^{I=0}(r)$ is expected to be sensitive to the change of the pion mass.
It is indeed the case, as shown in Fig.~\ref{fig:tensor_mass}: A magnitude of $V_T^{I=0}(r)$ becomes larger as the pion mass decreases\cite{Ishii:2010th}. 
\begin{figure}
\resizebox{0.5\textwidth}{!}{%
  \includegraphics{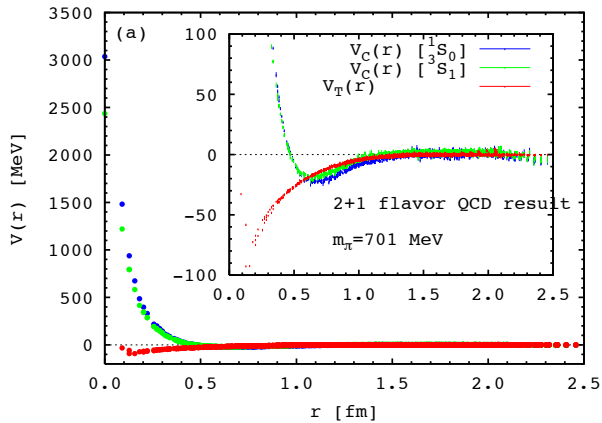}
}
\caption{Central and tensor potentials in 2+1 flavor QCD at $m_{\pi} \simeq 701$ MeV by the improved method. Taken from Ref.~\cite{Ishii:2010th}.}
\label{fig:full}       
\end{figure}

\begin{figure}
\resizebox{0.5\textwidth}{!}{%
  \includegraphics{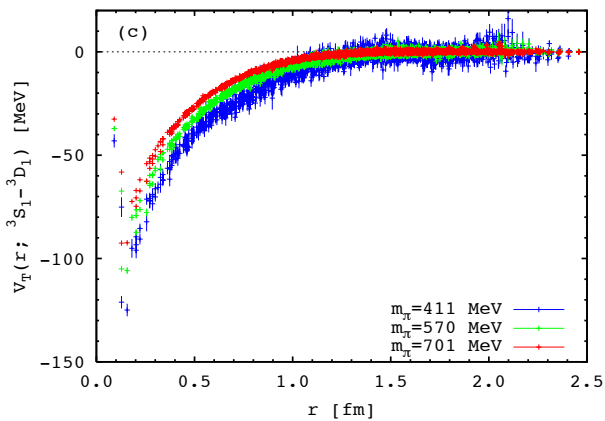}
}
\caption{Tensor potentials in 2+1 flavor QCD as a function of $r$ at $m_{\pi} \simeq 701$ MeV (red), 570 MeV (green) and 411 MeV (blue).  Taken from Ref.~\cite{Ishii:2010th}. }
\label{fig:tensor_mass}       
\end{figure}

\subsection{Convergence of the velocity expansion}
\begin{figure}
\resizebox{0.5\textwidth}{!}{%
  \includegraphics{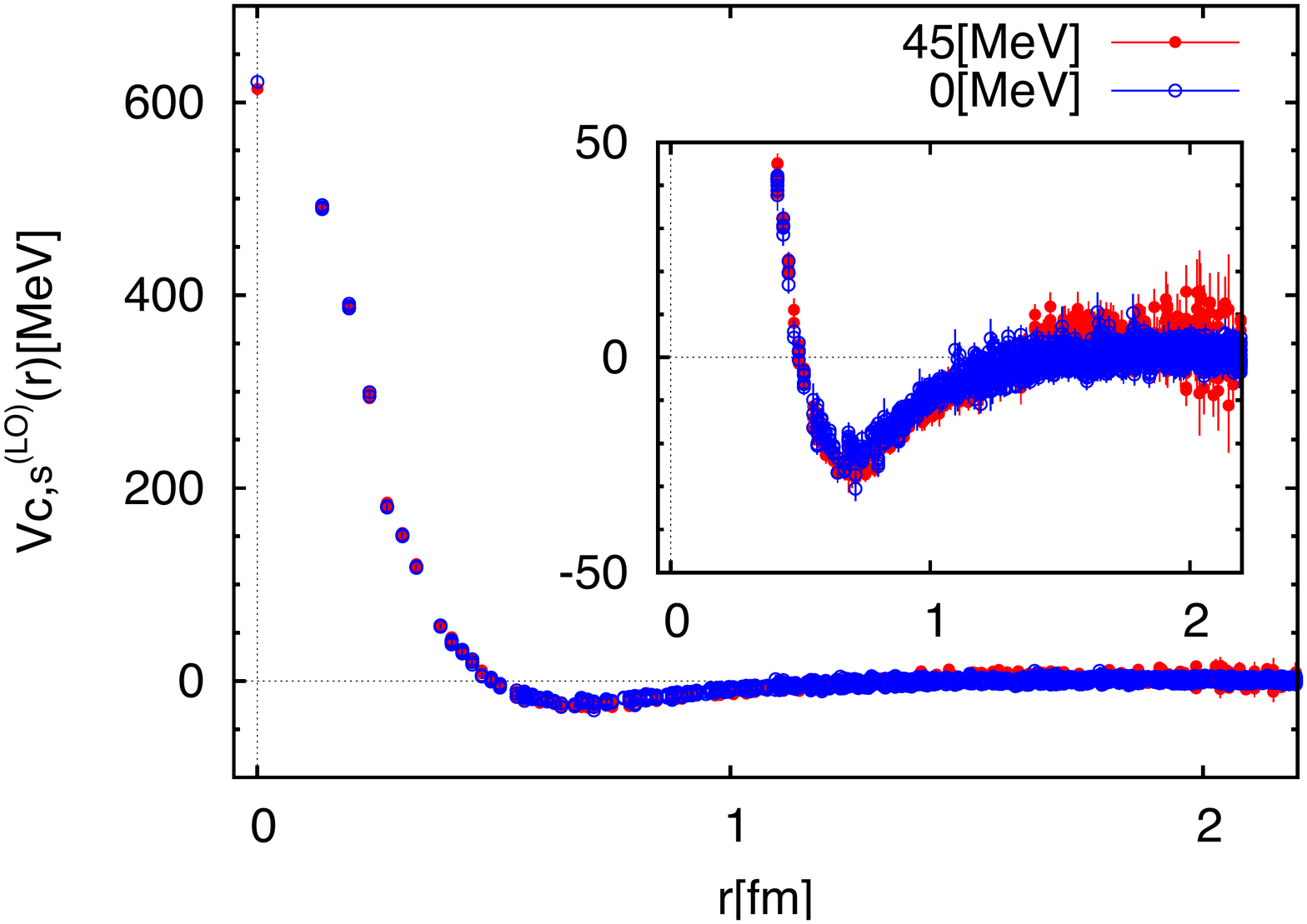}
}
\caption{The isospin-triplet (spin-singlet) central potential $V_C^{I=1}(r)$ obtained from the orbital $A_1^+$ representation at $E_k\simeq 45$ MeV (red) and at $E\simeq 0$ (blue) in quenched QCD at $m_{\pi} \simeq 529$ MeV. Taken from Ref.~\cite{Murano:2011nz}.}
\label{fig:E-dep_1S0}       
\end{figure}

\begin{figure}
\resizebox{0.5\textwidth}{!}{%
  \includegraphics{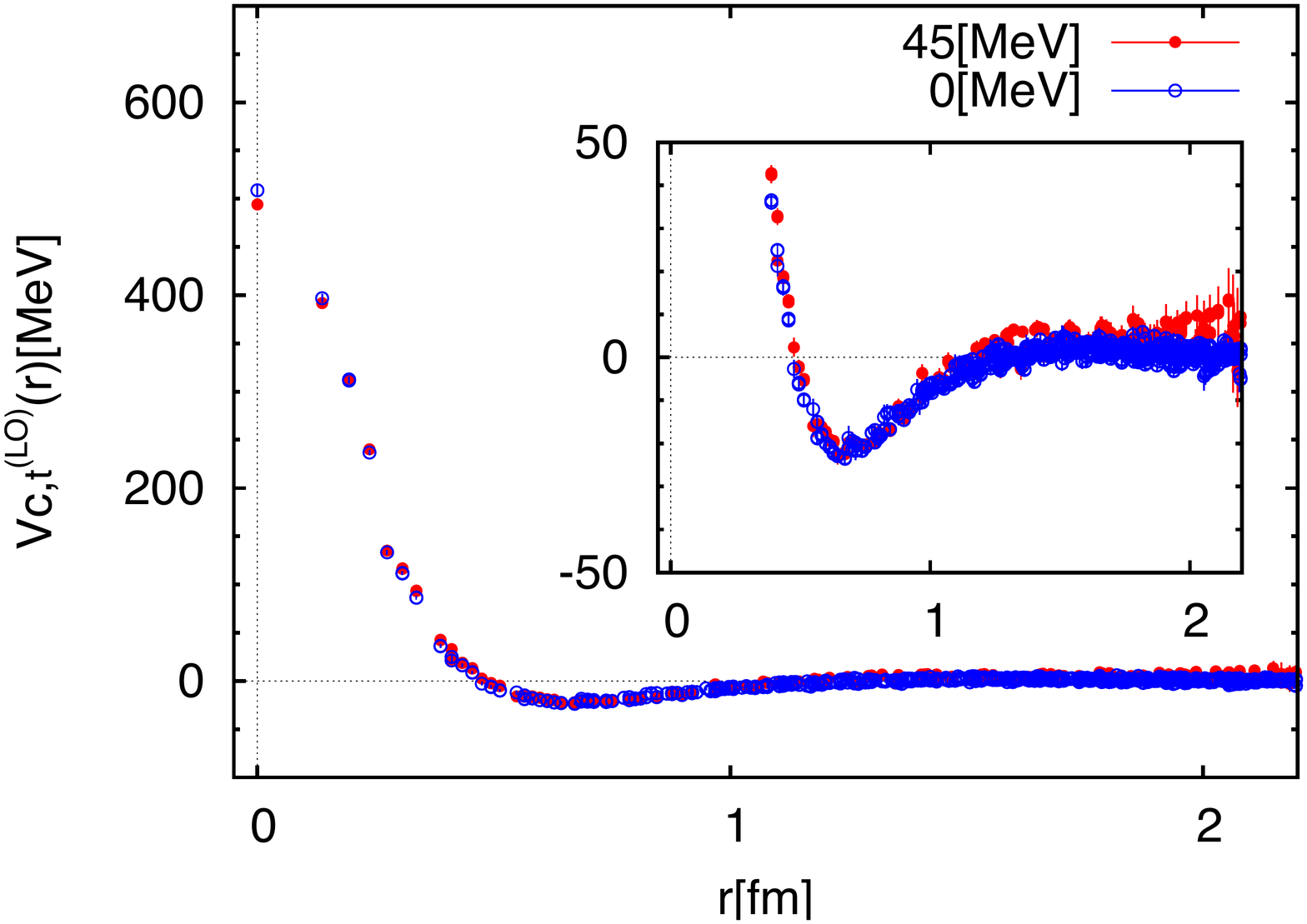}
}
\caption{The isospin-singlet (spin-triplet) central potential $V_C^{I=0}(r)$ obtained from the orbital $A_1^+-T_2^+$ coupled channel at $E_k\simeq 45$ MeV (red) and at $E\simeq 0$ (blue) in quenched QCD at $m_{\pi} \simeq 529$ MeV.  Taken from Ref.~\cite{Murano:2011nz}.}
\label{fig:E-dep_3S1}       
\end{figure}

\begin{figure}
\resizebox{0.5\textwidth}{!}{%
  \includegraphics{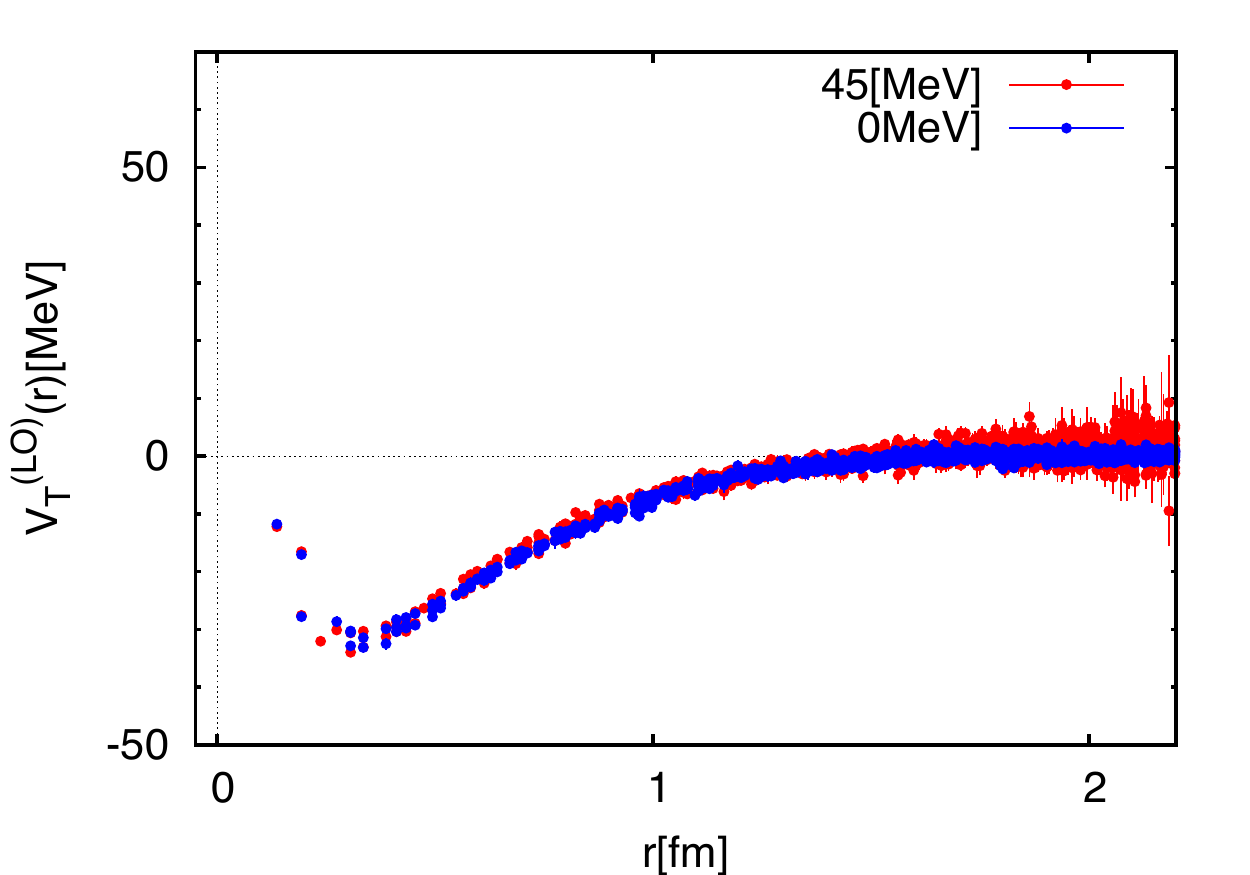}
}
\caption{The tensor potential $V_T^{I=0}(r)$ obtained from the orbital $A_1^+-T_2^+$ coupled channel at $E_k\simeq 45$ MeV (red) and at $E\simeq 0$ (blue) in quenched QCD at $m_{\pi} \simeq 529$ MeV.  Taken from Ref.~\cite{Murano:2011nz}.}
\label{fig:E-dep_tensor}       
\end{figure}
In this subsection, we discuss the convergence of the velocity expansion in eq.~(\ref{eq:V_expansion}).
If the non-locality of the potential were absent, the LO potential would be exact at all energies (below the inelastic threshold). Therefore, we can estimate a size of higher order terms in the velocity expansion,
by considering the energy dependence of the local potential at the LO.

Potentials shown so far are extracted with the periodic boundary condition in spatial directions for quark fields,
which leads to the almost zero kinetic energy $E_k$ for the ground state.
With  the anti-periodic boundary condition in spatial directions, on the other hand, we can significantly increase the ground state energy.  
To study the energy dependence of the LO potentials,
we thus  calculate the LO potentials at two different energies, $E_k\simeq 0$ MeV (periodic b.c.) and 45 MeV (anti-periodic b.c.),
in quenched QCD at $m_\pi \simeq 529$ MeV and $L\simeq 4.4$ fm,
by using the standard extraction of the potentials with the ground state saturation\cite{Murano:2011nz}.

With the anti-periodic boundary condition, 4 different momentum-wall sources, defined by
\beqa
q_f^{\rm wall}(t_0) &\equiv & \sum_{\bxs} q(\bx,t_0) f(\bx), 
\eeqa
are employed, where $f(\bx) = \cos( (\pm x\pm y + z)\pi/L)$, whereas $f(\bx)=1$ corresponds to the standard wall source used with the periodic boundary condition.
The momentum-wall sources generate the $L=T_2^+$ state  in addition to the $L=A_1^+$ state.

Fig.~\ref{fig:E-dep_1S0} compares the isospin-triplet (spin-singlet) central potential $V_C^{I=1}(r)$ obtained from $A_1^+$ state at $E_k\simeq 45$ MeV (red) with that at $E_k\simeq 0$ MeV (blue), while 
comparisons are made for  the isospin-singlet (spin-triplet) central potential $V_C^{I=0}(r)$  and  tensor potential $V_T^{I=0}(r)$  in Fig.~\ref{fig:E-dep_3S1} and Fig.~\ref{fig:E-dep_tensor}, respectively~\cite{Murano:2011nz}. 
They are obtained from the orbital $A_1^+-T_2^+$ coupled channel via eqs.~(\ref{eq:V3S1}) and (\ref{eq:VT}).
As seen from these figures, good agreements between two energies  for all three cases indicate that higher order contributions in the velocity expansion  are small in the energy region between 0 MeV and 45 MeV.
This means that the local potentials obtained at $E\simeq 0$ can be used to describe $NN$ scattering phase shifts in both isospin-triplet and -singlet channels for the energy up to 45 MeV, at this pion mass in quenched QCD.

\begin{figure}
\resizebox{0.5\textwidth}{!}{%
  \includegraphics{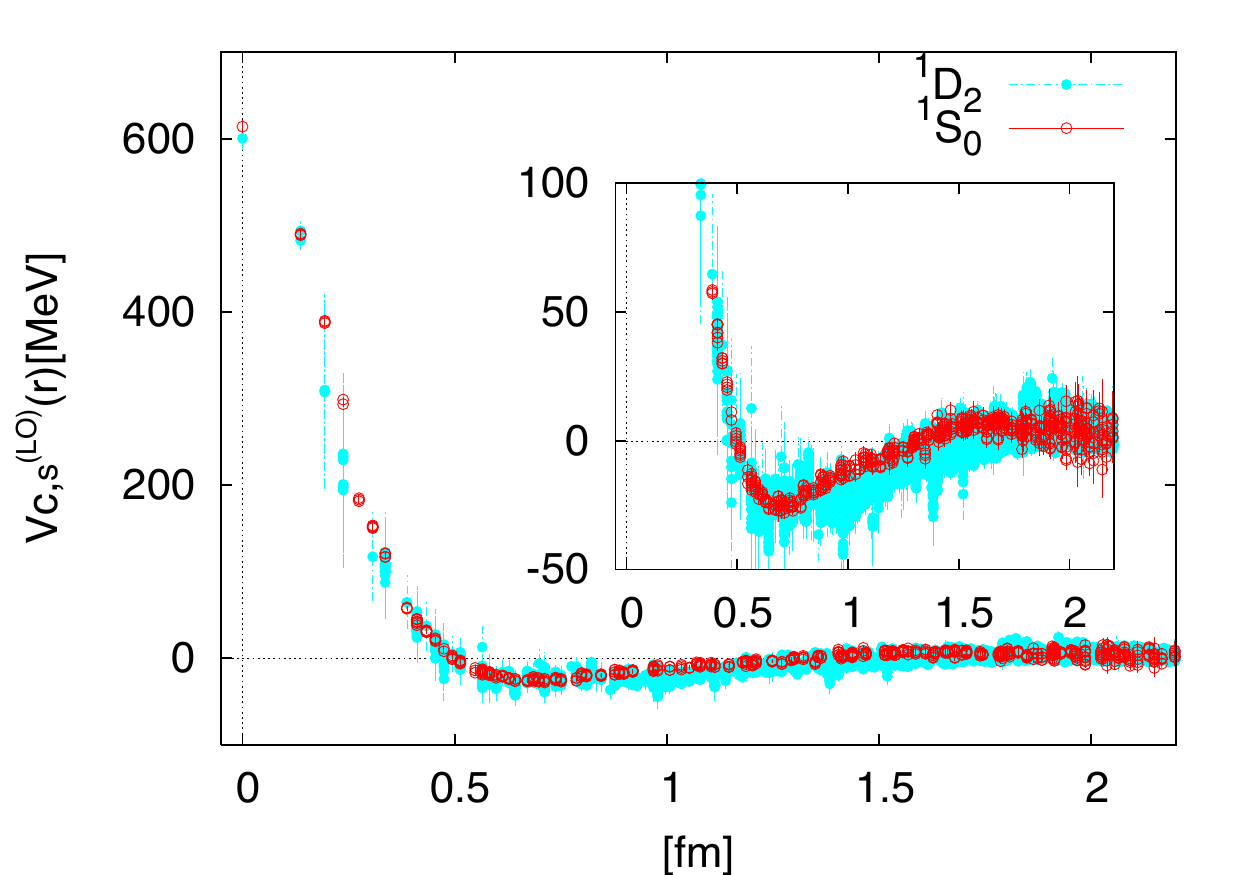}
}
\caption{The isospin-triplet (spin-singlet) central potential $V_C^{I=1}(r)$  at $E_k\simeq 45$ MeV, obtained from the orbital $A_1^+$ representation  (red) and from the $T_2^+$ representation (cyanogen), in quenched QCD at $m_{\pi} \simeq 529$ MeV.  Taken from Ref.~\cite{Murano:2011nz}.}
\label{fig:LL-dep}       
\end{figure}
Higher order contributions in the velocity expansion of the non-local potential may  also become manifest in the orbital angular momentum dependence of the local potential, since the orbital angular momentum $\bL = \br\times \bp$ contains one derivative.
Fig.~\ref{fig:LL-dep} compares the isospin-triplet (spin-singlet) potential $V_C^{I=1}(r)$ obtained from the $L=A_1^+$ state (red), whose main component has $L=0$, with the one from the $L=T_2^+$ state (cyanogen), which mainly has the $L=2$ component~\cite{Murano:2011nz}.  In both cases, the local potential is determined at the same energy, $E_k\simeq 45$,  MeV, but at different orbital angular momentum. Although statistical errors are large for the $L=T_2^+$ case, an agreement between the two is observed, suggesting that the $L$ dependence is also small. 

Comparisons in the above tell us that both energy and orbital angular momentum dependences for local LO potentials are almost invisible within statistical errors. We therefore conclude that contributions from higher order terms in the velocity expansion are small, so that  these LO potential obtained at $E\simeq 0$ and $L=0$ are good approximations for the non-local potentials at least up to the energy $E_k\simeq 45$ MeV for both $L=0$ and 2.  
  
\section{Conclusions and discussions}
We review  the HAL QCD method, recently proposed to extract hadronic interactions via lattice QCD simulations. We particularly focus on the methodology of this approach such as the strategy,  the theoretical foundation and numerical techniques, of the HAL QCD method.  

The equal-time NBS wave function for two nucleons plays a major role in the HAL QCD method, 
since it is proven that the NBS wave function encodes the $NN$ scattering phase shift in its asymptotic behavior at large separation. We therefore define the non-local potential, which can be constructed to be energy-independent, through the Schr\"odinger equation.  By construction, this non-local but energy-independent potential gives the correct phase shift  for the $NN$ scattering in QCD at all energies below the inelastic threshold.
In practice, we expand the non-local potential in terms of velocities and truncate this velocity expansion at the lowest few orders. Once the (approximated) potential is obtained in the velocity expansion, we can calculate the $NN$ phase shift approximately at all energies below the inelastic threshold, by solving the Schr\"odinger equation with this potential. 

In lattice QCD simulations, as in the case of the hadron mass extraction from a 2-pt correlation function, 
the NBS wave function can be extracted by the ground state saturation for the $NN$ 4-pt correlation functions, 
which however requires the large time separation and thus causes a large statistical fluctuation.
To overcome this difficulty, we have proposed an improve method which directly gives  the potential from the $NN$ 4-pt correlation functions without the ground state saturation.
We have shown in (2+1)-flavor  QCD simulations that the improved method works well to determine the isospin-triplet potential $V_C^{I=1}(r)$ at the LO of the velocity expansion, which reproduces a structure of the phenomenological potentials such as the  repulsive core at short distance surrounded by the attractive well at medium and long distances.
We have also shown the $NN$ scattering phase shift calculated with this potential as well as the isospin-singlet potentials $V_C^{I=0}(r)$ and $V_T^{I=0}(r)$ at the LO. 
We have estimated possible contributions from higher order terms in the velocity expansion, by studying the energy dependence as well as the $L$ dependence of the LO potentials in quenched QCD and have found that
such contributions are rather small at low energy and small $L$. 

The HAL QCD method is general and can be applied to other cases, which will be mentioned below with some references.

So far $NN$ potentials in our calculations are restricted to the parity even sector.  We recently extend our study to the parity odd sector including the $\bL\cdot\bS$ potential, which appears at the next-to-leading order of the velocity expansion\cite{Murano:2011aa}. 
In the HAL QCD method, the three nuclear force can also be investigated and an indication of the short-distance repulsion is found in 2-flavor QCD simulations with the heavy pion mass\cite{Doi:2010yh,Doi:2011gq}.

The HAL QCD method for the $NN$ potentials can be easily applied to potentials between other baryons.  The $N\Xi$ potential was calculated
in quenched QCD simulations, as the first attempt to study  nucleon-hyperon interactions\cite{Nemura:2008sp}, and soon after the $N\Lambda$ potential 
 has been calculated in both quenched and full QCD simulations\cite{Nemura:2009kc}.
As more general cases, interactions between octet baryons are investigated  in the flavor SU(3) limit, where up, down and strange quark masses are all degenerate\cite{Inoue:2010hs,Inoue:2011ai}.
Among 6 independent potentials corresponding to irreducible representations of the flavor SU(3) group,
the flavor singlet potential is found to be attractive enough at all distances  to have one bound state\cite{Inoue:2010es}, which corresponds to the H-dibaryon,  predicted in Ref.~\cite{Jaffe:1976yi}.
In order to analyze the property of the H-dibaryon  in  the real world where the strange quark is much heavier that the up and down quarks, the generalization of the HAL QCD method to the coupled channel potentials  is required\cite{Ishii:2011tq,Aoki:2011gt}, with which the baryon-baryon potentials in the $S=-2$ sector are investigated in (2+1)-flavor  QCD simulations\cite{Sasaki:2010bi,Sasaki:2011zz}, where $S$ represents the strangeness.

There are a few studies on meson-baryon interactions by the HAL QCD method such as
$KN$\cite{Ikeda:2010sg,Ikeda:2011qm} and charmonium and nucleon\cite{Kawanai:2010ev}. 

Short distance behaviors of the potentials  defined in the HAL QCD scheme can be investigated analytically by the operator product expansion and the renormalization group in the perturbative QCD, for the $NN$\cite{Aoki:2009pi,Aoki:2010kx},
the baryon-baryon\cite{Aoki:2010uz}, the $3N$\cite{Aoki:2011aa}, and  the 3 baryons\cite{Aoki:2012xa}.
See Ref.~\cite{Aoki:2013zj} for a review on these results.   

It is shown recently that the energy independent non-local coupled channel potentials exist even above the inelastic threshold\cite{Aoki:2012bb}. This result opens a possibility that the hadronic interaction can be extracted in lattice QCD at all energies without theoretical restrictions..

Finally, needless to say, the next step must be to calculate the $NN$ potential at physical pion mass.
It is a great challenge for lattice QCD  to show that the deuteron indeed bounds while  di-neutron does not     
at the physical point.  

%
%
\section*{Acknowledgement}
I would like to thank all members of the HAL QCD collaboration for useful discussions during researches performed together.
This work is supported in part by the Grants-in-Aid for Scientific Research on Innovative Areas (No.2004: 20105001, 20105003) and SPIRE (Strategic Program for Innovative Research). 

\section*{Appendix: Asymptotic behavior of the NBS wave function}
\label{sec:app1}
While the asymptotic behavior of the NBS wave function at large $r$ for the elastic $\pi\pi$ case has been shown in Refs.~\cite{Lin:2001ek,Aoki:2005uf} and  extended to the elastic $NN$ case in Ref.~\cite{Ishizuka2009a},
 we give a different derivation for it  in this appendix, using the Lippmann-Schwinger equation (\ref{eq:LSE}).
For simplicity, we consider  the case of the scalar fields.

The unitarity of the S-matrix implies
\beqa
T^\dagger - T &=& i T^\dagger T,
\eeqa
which can be solved for the two particle scattering  as
\beqa
T \equiv \frac{1}{2\pi} T(\bq_1, \bq_2) &=&  \frac{1}{2\pi}\sum_{L,M} T_L(q,q) Y_{LM}(\Omega_{\bqs_1}) \overline{Y_{LM}(\Omega_{\bqs_2})}, \nn \\
T_L(q,q) &=& -\frac{8}{qE_q} e^{i\delta_L(q)} \sin\delta_L(q), 
\label{eq:unitarity}
\eeqa
where $q=\vert\bq_1\vert = \vert \bq_2\vert$, $Y_{LM}$ is the spherical harmonic function, $\Omega_{\bqs}$ is the solid angle of the vector $\bq$, and $\delta_L(q)$ is the scattering phase shift for the partial wave with the angular momentum $L$ at energy $E_q=2\sqrt{q^2+m^2}$ with  the mass of the scalar particle $m$.

The equal-time NBS wave function for two scalar particle is defined by
\beqa
\Psi_{\bks}(\bx) &=& {}_{\rm in}\langle 0 \vert \varphi^2(\bx,0)\vert \bk \rangle_{\rm in},\\
\varphi^2(\bx,0) &\equiv& T\{\varphi_1(\br+\bx,0) \varphi_2(\br,0)\} \nn
\eeqa
where,  for simplicity, we assume that two scalar fields $\varphi_1$ and $\varphi_2$ have the same mass $m$.  From the Lippmann-Schwinger equation (\ref{eq:LSE}), we have
\beqa
\vert 0 \rangle_{\rm in} &=& \vert 0\rangle_0 + \int d\bq\, \frac{\vert\bq \rangle_0 T_{\bqs 0}}{E_0-E_q +i\varepsilon}
\label{eq:vacuum}
\eeqa
for the vacuum instate. The contribution of eq.~(\ref{eq:vacuum})  to the NBS wave function at large $r=\vert\bx\vert$ amounts to
\beqa
{}_{\rm in}\langle 0\vert \varphi^2(\bx,0)\vert \bk \rangle_0 &\simeq& \frac{1}{Z_k} {}_0\langle 0\vert \varphi^2(\bx,0)\vert\bk \rangle_0, 
\eeqa
where $Z_k$ is the normalization factor whose deviation from unity comes from the off-shell $T$-matrix $T_{\bqs 0}$.  Using this, the NBS wave function becomes
\beqa
\Psi_{\bks}(\bx) &=& \frac{1}{Z_k} {}_0\langle 0\vert \varphi^2(\bx,0)\vert\bk \rangle_0\nn \\
&+&\int d\bq\, \frac{1}{Z_q} \frac{{}_0\langle 0\vert \varphi^2(\bx,0)\vert\bq \rangle_0 T_{\bqs\bks}}{E_{\bks} - E_{\bqs} +i\varepsilon}.
\label{eq:LS_NBS}
\eeqa
Inserting the expression that
\beqa
{}_0\langle 0\vert \varphi^2(\bx,0)\vert \bk \rangle_0 &=& \frac{1}{(2\pi)^3 2 E_k}e^{i \bks\cdot \bxs} ,
\eeqa
with $\vert \bk \rangle_0 \equiv a_1^\dagger(\bk) a_2^\dagger(-\bk)\vert 0 \rangle_0 $, into eq.~(\ref{eq:LS_NBS}), we have
\beqa
\Psi_{\bks}(\bx) &=&\frac{1}{2E_k Z_k}\Bigl[\frac{e^{i\bks\cdot\bxs}}{(2\pi)^3} \nn \\
&+&\int \frac{d^3q}{(2\pi)^3}
\frac{Z_kE_k}{Z_qE_q}\frac{e^{i\bqs\cdot\bxs} T(\bq,\bk)}{4\pi(E_k-E_q+i\epsilon)} \Bigr].
\eeqa

With expressions that
\beqa
e^{i\bks\cdot\bxs} &=& 4\pi \sum_{L,M} i^L j_L(kr) Y_{LM}(\Omega_{\bxs})\overline{Y_{LM}(\Omega_{\bks})},\\
\Psi_{\bks}(\bx) &=& \sum_{L,M} i^l \Psi_L(r,k) Y_{LM}(\Omega_{\bxs})\overline{Y_{LM}(\Omega_{\bks})},
\eeqa
where $j_L(x)$ is the spherical Bessel function of the first kind, 
we obtain
\beqa
\Psi_L(r,k) &=& \frac{4\pi}{(2\pi)^3 2E_k Z_k}\Bigl[ j_L(kr) \nn\\
&+&\int_0^\infty \frac{q^2dq}{2\pi}\frac{Z_kE_k}{Z_qE_q} \frac{j_L(qr) T_L(q,k)}{2(E_k-E_q+i\varepsilon)}\Bigr].
\label{eq:NBS_l}
\eeqa
Under the assumption that $T_L(k,q)$ does not have any poles in the positive real axis
 for $E_k$ below the inelastic threshold, we perform the $q$ integral using the formula
 \beqa
 \int dq \frac{j_L(qr)}{k^2-q^2+i\varepsilon} F_L(q) &\simeq& -\frac{\pi}{2k} F_L(k) \left[h_L(kr) + i j_L(kr)\right]
 \nn
 \eeqa
for $r \gg 1$, where $F_L(q)$ does not have any poles in the positive real axis and satisfies $F_L(-q) = (-1)^L F_L(q)$, and $n_L(x)$ is the the spherical Bessel function of the second kind. 
After the $q$ integral, the second term in eq.~(\ref{eq:NBS_l}) becomes
\beqa
&-&[n_L(kr) + ij_L(kr)] \frac{k E_k}{8} T_L(k,k)  \nn \\
&=&[n_L(kr) + ij_L(kr)]
\times e^{i\delta_L(k)}\sin \delta_L(k),
\eeqa
where the unitarity constraint (\ref{eq:unitarity}) for $T_L(k,k)$ is used.
We then finally obtain
\beqa
\Psi_L(r,k) &\simeq & 
C_L(k)
\left[j_L(kr)\cos\delta_L(k)+n_L(kr)\sin\delta_L(k)\right] \nn \\
&\simeq &
C_L(k) \frac{\sin(kr-L\pi/2+\delta_L(k))}{kr}
\eeqa
for $r\gg 1$, where asymptotic behaviors that $j_L(x)\simeq \sin(x-L\pi/2)/x$ and  $n_L(x)\simeq \cos(x-L\pi/2)/x$ are used, and the constant $C_L(k)$ is given by
\beqa
C_L(k) &=& \frac{4\pi e^{i\delta_L(k)}}{(2\pi)^32E_k Z_k} .
\eeqa
The phase appeared in the $T$-matrix , $\delta_L(k)$, can be interpreted as the scattering phase shift of the NBS wave function. 

%

\begin{thebibliography}{}
%
%
\bibitem{Fodor:2012gf}
  Z.~Fodor and C.~Hoelbling,
  Rev.\ Mod.\ Phys.\  \textbf{84} (2012) 449.
  
\bibitem{Luscher:1990ux}
M.~L\"{u}scher, Nucl.\ Phys.\  \textbf{B354}  (1991) 531.

\bibitem{Orginos:2011zz}
  K.~Orginos,
  PoS \textbf{LATTICE2011} (2011) 016.
  
\bibitem{Ishii:2006ec}
  N.~Ishii, S.~Aoki and T.~Hatsuda,
Phys.\ Rev.\ Lett.\ \textbf{99} (200) 022001.

\bibitem{Aoki:2008hh}
  S.~Aoki, T.~Hatsuda and N.~Ishii,
Comput.\ Sci.\ Dis.\   \textbf{1} (2008)  015009.

\bibitem{Aoki:2009ji}
  S.~Aoki, T.~Hatsuda and N.~Ishii,
Prog.\ Theor.\ Phys.\ \textbf{123} (2010)  89.

\bibitem{Nemura:2008sp}
  H.~Nemura, N.~Ishii, S.~Aoki and T.~Hatsuda,
Phys.\ Lett.\  \textbf{B673} (2009) 136.  
  
\bibitem{Nemura:2009kc}
  H.~Nemura, N.~Ishii, S.~Aoki and T.~Hatsuda  [PACS-CS Collaboration],
  PoS \textbf{LATTICE2008} (2008) 156.

\bibitem{Inoue:2010hs}
  T.~Inoue {\it et al.}  [HAL QCD collaboration],
Prog.\ Theor.\ Phys. \textbf{124} (2010) 591.

\bibitem{Inoue:2010es}
  T.~Inoue {\it et al.}  [HAL QCD Collaboration],
Phys.\ Rev.\  Lett.\ \textbf{106} (2011) 162002. 

\bibitem{Inoue:2011ai} 
  T.~Inoue {\it et al.}  [HAL QCD Collaboration],
  Nucl.\ Phys.\  \textbf{A881} (2012) 28. 

  \bibitem{Ikeda:2010sg}
  Y.~Ikeda {\it et al.},
PoS \textbf{LATTICE2010} (2010) 143.  

\bibitem{Ikeda:2011qm}
  Y.~Ikeda [HAL QCD Collaboration],
  PoS \textbf{LATTICE  2011} (2011) 159
  [arXiv:1111.2663 [hep-lat]].

\bibitem{Kawanai:2010ev}
  T.~Kawanai and S.~Sasaki,
  Phys.\ Rev. \ \textbf{D82} (2010) 091501.

\bibitem{Doi:2010yh}
  T.~Doi  for HAL QCD Collaboration,
  PoS\ \textbf{LATTICE2010} (2010) 136.  

\bibitem{Doi:2011gq} 
  T.~Doi {\it et al.}  [HAL QCD Collaboration],
  Prog.\ Theor.\ Phys.\  \textbf{127} (2012) 723. 
  
\bibitem{Aoki:2011ep} 
  S.~Aoki for HAL QCD Collaboration,
  Prog.\ Part.\ Nucl.\ Phys.\  \textbf{66} (2011) 687.

\bibitem{Aoki:2012tk} 
  S.~Aoki {\it et al.}  [HAL QCD Collaboration],
  Prog.\ Theor.\ Exp.\ Phys. \textbf{2012}  (2012) 01A106. 

 \bibitem{text_weinberg}
 S.~Weinberg,  \textit{The Quantum Theory of Fields, Volume I Foundations} (Cambridge University Press, Cambridge, United Kingdom, 1999) p.155, Chapter 3. 

\bibitem{Balog:2001wv} 
  J.~Balog, M.~Niedermaier, F.~Niedermayer, A.~Patrascioiu, E.~Seiler and P.~Weisz,
  Nucl.\ Phys.\  \textbf{B618}  (2001) 315.

\bibitem{Ishizuka2009a}
N.~Ishizuka,  PoS\  \textbf{LAT2009} (2009) 119.

\bibitem{TW67}
R. Tamagaki and W. Watari, 
Prog.\ Theor.\ Phys.\  Suppl.\ \textbf{39} (1967) 23.

\bibitem{okubo1958}
S.~Okubo and R.~E.~Marshak, 
Ann.\ Phys.\  \textbf{4}  (1958) 166. 

\bibitem{Nishijima}
K. ~Nishijima, Phys.\ Rev.\  \textbf{111} (1958) 153.

\bibitem{Zimmermann}
W.~Zimmermann, Nuovo Cim.\  \textbf{10} (1958) 597.

\bibitem{Haag}
R.~Haag, Phys.\ Rev.\ \textbf{112}  (1958) 669.

\bibitem{Parisi:1983ae} 
  G.~Parisi,
  Phys.\ Rept.\  \textbf{103} (1984) 203.
  
\bibitem{Lepage:1989hd}
  G.~P.~Lepage,
  in {\it From Actions to Answers: Proceedings of the TASI 1989},
  edited by T.~Degrand and D.~Toussaint
  (World Scientific, Singapore, 1990).

\bibitem{HALQCD:2012aa}
  N.~Ishii {\it et al.}  [HAL QCD Collaboration],
  Phys.\ Lett.\  \textbf{B712}  (2012) 437. 

\bibitem{Kuramashi:1995sc}
  Y.~Kuramashi,
  Prog.\ Theor.\ Phys.\ Suppl.\  {\bf 122} (1996) 153.

\bibitem{Aoki:2008sm}
  S.~Aoki {\it et al.}  [PACS-CS Collaboration],
  Phys.\ Rev.\   \textbf{D79} (2009) 034503.

\bibitem{Machleidt:2000ge}
  R.~Machleidt,
  Phys.\ Rev.\  \textbf{C63} (2001) 024001.

\bibitem{Stoks:1994wp}
  V.~G.~J.~Stoks, R.~A.~M.~Klomp, C.~P.~F.~Terheggen and J.~J.~de Swart,
  Phy.\ Rev.\  \textbf{C49} (1994) 2950.

\bibitem{Wiringa:1994wb}
  R.~B.~Wiringa, V.~G.~J.~Stoks and R.~Schiavilla,
  Phy.\ Rev.\  \textbf{C51} (1995) 38.
 
 \bibitem{nature}
 In {\it Research Highlights 2007}, Nature \textbf{450} (2007) 1130.
 
\bibitem{nn-online}
  http://www.nn-online.org/

\bibitem{Ishii:2010th}
  N.~Ishii [PACS-CS and HAL-QCD Collaborations],
  PoS  \textbf{LAT2009} (2009) 019.

\bibitem{Murano:2011nz}
  K.~Murano, N.~Ishii, S.~Aoki and T.~Hatsuda,
  Prog.\ Theor.\ Phys.\  \textbf{125} (2011) 1225.

\bibitem{Murano:2011aa}
  K.~Murano [HALQCD Collaboration],
  PoS \textbf{LATTICE2011} (2011) 319.

\bibitem{Jaffe:1976yi}
  R.~L.~Jaffe,  
  Phys.\ Rev.\ Lett.\  \textbf{38} (1977) 195
  [Erratum-ibid.\  \textbf{38} (1977) 617].

\bibitem{Ishii:2011tq}
N.~Ishii, for HAL QCD Collaboration,
PoS \textbf{LATTICE2010} (2010) 145.    
  
\bibitem{Aoki:2011gt}
  S.~Aoki {\it et al.}  [HAL QCD Collaboration],
Proc.\ Japan Acad.\  \textbf{B87}  (2011) 509. 

\bibitem{Sasaki:2010bi}
  K.~Sasaki [HAL QCD Collaboration],
  PoS \textbf{LATTICE2010} (2010) 157.
    
 \bibitem{Sasaki:2011zz}
  K.~Sasaki [HAL QCD Collaboration],
  PoS \textbf{LATTICE2011} (2011) 173.

\bibitem{Aoki:2009pi}
  S.~Aoki, J.~Balog and P.~Weisz,
  PoS \textbf{LATTICE 2009} (2009) 132.

\bibitem{Aoki:2010kx}
  S.~Aoki, J.~Balog and P.~Weisz,
  JHEP \textbf{1005} (2010) 008.

 \bibitem{Aoki:2010uz}
  S.~Aoki, J.~Balog and P.~Weisz,
  JHEP \textbf{1009} (2010) 083.

\bibitem{Aoki:2011aa}
  S.~Aoki, J.~Balog and P.~Weisz,
  New J.\ Phys.\  \textbf{14} (2012) 043046.
  
\bibitem{Aoki:2012xa}
  S.~Aoki, J.~Balog and P.~Weisz,
  Prog.\  Theor.\  Phys.\  \textbf{128} (2012) 1269.

\bibitem{Aoki:2013zj}
  S.~Aoki, J.~Balog, T.~Doi, T.~Inoue and P.~Weisz,
  Int.\ J.\ Mod.\ Phys.\ E {\bf 22} (2013) 1330012.

\bibitem{Aoki:2012bb}
  S.~Aoki, B.~Charron, T.~Doi, T.~Hatsuda, T.~Inoue and N.~Ishii,
  Phys.\ Rev.\ D {\bf 87} (2013) 034512.
  
\bibitem{Lin:2001ek}
  C.~J.~D.~Lin, G.~Martinelli, C.~T.~Sachrajda and M.~Testa,
  Nucl.\ Phys.\  \textbf{B619} (2001) 467.

\bibitem{Aoki:2005uf} 
  S.~Aoki {\it et al.}  [CP-PACS Collaboration],
  Phys.\ Rev.\  \textbf{D71} (2005)  094504.

\end{thebibliography}
%

\end{document}